\documentclass{iau}
\usepackage{graphicx}

\title[JD 11.~~Pulsations in B supergiants] %% give here short title %%
{Pulsations in the late-type B supergiant star HD 202850\thanks{Based on observations acquired at the
  Dominion Astrophysical Observatory, Herzberg Institute of
  Astrophysics, National Research Council of Canada}}

\author[Sanja Tomi\'c, Michaela Kraus \& Mary E. Oksala]   %% give here short author list %%
{Sanja Tomi\'c$^1$, 
Michaela Kraus$^2$
 \and Mary Oksala$^2$}

\affiliation{$^1$Department of Astronomy, Faculty of Mathematics, University of Belgrade\\ Studentski trg 16,
11000 Belgrade, Serbia \\ email: {\tt sanja@sunstel.asu.cas.cz} \\[\affilskip]
$^2$Astronomick\'y \'ustav, Akademie v\v{e}d \v{C}esk\'e republiky\\ Fri\v{c}ova 298, 25165 Ond\v{r}ejov, Czech Republic}

\pubyear{2013}
\volume{301}  %% insert here IAU Symposium No.
\pagerange{1--2}
\jname{Precision Asteroseismology}
\editors{J.A. Guzik, W.J. Chaplin, G. Handler \& A. Pigulski, eds.}
\begin{document}

\maketitle

\begin{abstract}
HD 202850 is a late B-type supergiant. It is known that
photospheric lines of such stars vary. Due to
macroturbulence the lines are much wider than
expected. Macroturbulence has been linked to stellar
pulsations. It has been reported that there are several B
supergiants that undergo pulsations. In our previous work, we detected a pulsational period of
$1.59$ hours in this object from data taken with the
Ond\v{r}ejov 2-m telescope. We continued to investigate this
object and we took several time series with the DAO
1.2-m telescope. Our new data suggest that there may
be some additional pulsational periods in this star. We
present our new results in this poster.
%% add here a maximum of 10 keywords, to be taken form the file <Keywords.txt>
\end{abstract}

\firstsection % if your document starts with a section,
              % remove some space above using this command.
\section{Introduction}

B supergiants are very important for stellar and galactic evolution, as they enrich their evironments with chemically processed material via their line-driven winds. They show strong line profile variability. Their lines are wider than expected from their parameters. The excessive width is due to macroturbulence. Both line profile variability and macroturbulence are indications of stellar pulsations. However, so far only very few such supergiants were investigated to determine their pulsation periods.\\ 

\section{HD 202850 ($=\sigma$\,Cyg)}

HD 202850 is a late B-type supergiant star. Its stellar parameters (\cite{Markova}) are given in Table~\ref{param}. It has been classified as B9 Iab, and is located in the OB association Cyg OB 4 at a distance of \,$\approx$\,1 kpc. It falls out of any previously calculated instability domains (\cite{Siao}). In our previous work, we described the 1.59 h pulsation period we detected (see \cite{Kraus}).\\
 
\begin{table}[h]
\centering
\caption{Parameters of HD 202850}
\scriptsize{
\begin{tabular}{|c|c|c|c|c|c|c|}
\hline
$T_{\rm eff}$ & $\log L/L_{\odot}$ & $\log g$ & $R_{*}$ & $M$ & $v\sin i$ & $v_{\rm macro}$\\\hline
[K] & & & [$R_{\odot}$] & [$M_{\odot}$] & [km/s] & [km/s]\\
\hline 
$11000$&$4.59$&$1.87$&$54$& $8_{-3}^{+4}$ & $33\pm2$& $33\pm2$\\
\hline

\end{tabular}
}
\label{param}
\end{table}

\section{New preliminary results}

\begin{figure}[h]
% \vspace*{-2.0 cm}
\begin{center}
 \includegraphics[width=4in]{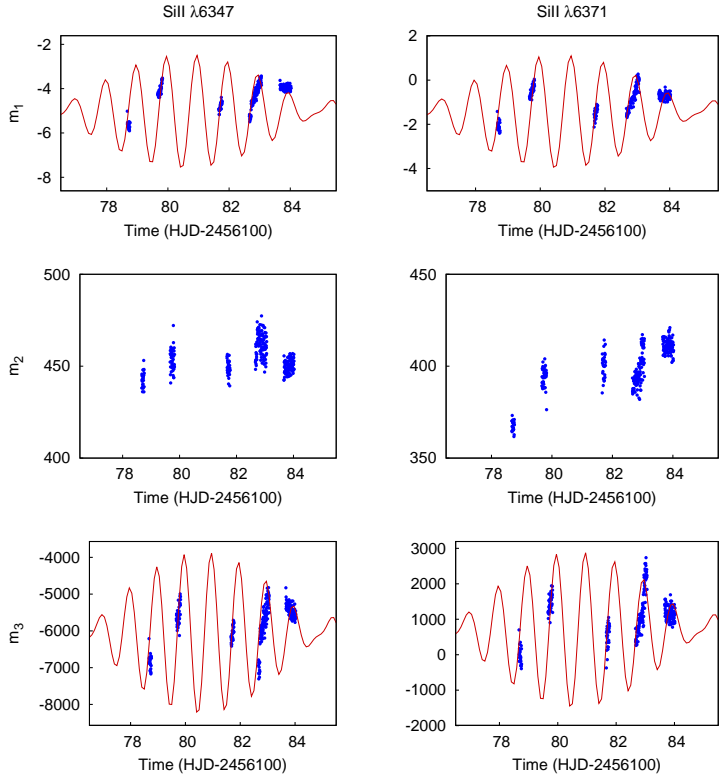} 
% \vspace*{-1.0 cm}
 \caption{First three small moments.  A combination of possible two periods is fit through the first and the third moment.}
   \label{fig1}
\end{center}
\end{figure}

In 2012 we took a new set of 294 spectra distributed over 5 nights with the DAO 1.2-m telescope. Exposures were five minutes long, with a signal-to-noise ratio between 150 and 250. The moment analysis showed variability in all three moments, and the first and the third moment seem to vary in phase. Due to high noise, the FFT analysis did not show any pronounced peaks, therefore the period(s) were estimated by fitting a combination of sine curves (Fig.~\ref{fig1}). We found two new possible periods (a $22.2$\,h period and a $25.2$\,h period).

\begin{acknowledgements}
S.T. acknowledges financial support from an IAU grant. M.K. and M.E.O. acknowledge financial support from GA\v{C}R under grant number P209/11/1198. The Astronomical Institute Ond\v{r}ejov is supported by the project RVO:67985815.
\end{acknowledgements}

\end{document}